\documentclass[10pt,a4paper]{article} 
\usepackage[dvips]{graphicx}
\usepackage{amsfonts,amsmath,amssymb}
\usepackage{hyperref}

\begin{document}
\title{Massive Jackiw-Rebbi Model}

\author{
F. Charmchi\footnote{Electronic address: f$\_$charmchi@sbu.ac.ir}~~and S.S. Gousheh\footnote{Electronic address: ss-gousheh@sbu.ac.ir}\\
 \small   Department of Physics, Shahid Beheshti University G.C., Evin, Tehran
19839, Iran}
\maketitle
\begin{abstract}
In this paper we analyze a generalized Jackiw-Rebbi (J-R) model in which a massive fermion 
is coupled to the kink of the $\lambda\phi^4$ model as a prescribed background field.
We solve this massive J-R model exactly and analytically and obtain the whole spectrum 
of the fermion, including the bound and continuum states.
The mass term of the fermion makes the potential of the decoupled second order Schrodinger-like 
equations asymmetric in a way that their asymptotic values at two spatial infinities are different.
Therefore, we encounter the unusual problem in which two kinds of continuum states are possible 
for the fermion: reflecting and scattering states.
We then show the energies of all the states as a function of the parameters of the kink, 
i.e.\;its value at spatial infinity ($\theta_0$) and its slope at $x=0$ ($\mu$).
The graph of the energies as a function of $\theta_0$, where the bound state energies and the 
two kinds of continuum states are depicted, shows peculiar features including an energy gap in 
the form of a triangle where no bound states exist.
That is the zero mode exists only for $\theta_0$ larger than a critical value $(\theta_0^{\textrm{c}})$.
This is in sharp contrast to the usual (massless) J-R model where the zero mode and hence the 
fermion number $\pm1/2$ for the ground state is ever present.
This also makes the origin of the zero mode very clear: It is formed from the union 
of the two threshold bound states at $\theta_0^{\textrm{c}}$, 
which is zero in the massless J-R model.
\end{abstract}

\section{Introduction}
In 1976 Jackiw and Rebbi \cite{jackiw-rebbi} introduced the important concept of the 
fractional fermion number of the solitons, considering two different fermion-soliton 
models, one of them in one and the other in three spatial dimensions.  
In both models, the key observations that lead to the fractional charge of the soliton 
are that the models possess charge conjugation symmetry
and also there is a nondegenerate zero-energy fermionic mode.
They showed that in the presence of the zero mode the prescribed 
soliton is a degenerate doublet carrying charge $\pm 1/2$.
In the ensuing decades there has been a vast number of works confirming and elaborating on the J-R finding.
This discovery has motivated much of the works on this subject and the concept of the vacuum polarization by 
background fields has been investigated in many branches of physics such as particle physics 
\cite{jackiw-rebbi,goldstone,mackenzie1,mackenzie2,dr1,dr2,jackiw,witten,rajaraman,jaffe,leila,dehghan}, 
cosmology \cite{cosm1,cosm2,cosm3,cosm4,cosm5}, condensed-matter physics \cite{cond1,cond2,cond3,cond4}, 
polymer physics \cite{poly1,poly2,poly3} and atomic physics \cite{atom1,atom2,atom3}.
\par
We now explain some of these works which are relevant to this paper.
In 1981 Goldstone and Wilczek \cite{goldstone} invented a powerful method, called the adiabatic method,
for calculating the vacuum polarization of fermions induced by the background solitons.
In this method the final topological background field, which is assumed to be slowly varying in space, 
is considered to be slowly evolving from the topologically trivial configuration. 
Using their method, they investigated some models which lack the symmetry in the 
energy spectrum of the fermion
and showed that the fermion number of the vacuum can be any real number and not just $\pm 1/2$.
Later on this method was generalized by MacKenzie and Wilczek \cite{mackenzie1,mackenzie2}.
In their method the requirement of the slow spatial variation of the background field was lifted and 
therefore they could consider models including solitons with arbitrary variations in the space.
Using their method, they concluded that sharply varying solitons can never polarize the vacuum.  
Following these works, some authors used these methods to investigate the vacuum polarization for different models.
In one of these papers, the authors \cite{dr1} studied an exactly solvable model in which a 
fermion is coupled to a background field with two adjustable parameters. 
By varying these parameters, one can have different topological background 
fields with different topological charges and scale of variation.
Using this simple model, they were able to explore the effect of the scale of variations of the solitons on the vacuum polarization. 
\par
In the J-R model there is no explicit mass term for the Fermi field and the zero mode 
is always present, regardless of the values of the parameters of the model.
These parameters are the Yukawa coupling constant, denoted by g, the values of the background 
field at spatial infinity, denoted by $\theta_0$, and its slope at zero, denoted by $\mu$.
In a previous paper we presented exact solutions for the J-R model and showed explicitly 
that there is a dynamically generated mass $M_0=g\theta_0$ \cite{farid}.
We also reasoned that as $\theta_0$ increases and a mass gap appears in the spectrum, the 
two threshold bound states which separated the continua at $\theta_0=0$, join to form the 
ever present self-charge-conjugate nondegenerate zero mode.
In this paper we generalize the J-R model by adding an explicit mass term for the Fermi field, 
denoted by $M$, and solving the dynamical equations exactly, we find that the system possesses 
some unusual properties.
In particular the potentials appearing in the two Schrodinger-like equations obtained from 
decoupling the Dirac equation have unequal values at $x\to\pm\infty$.
Therefore, we have, in addition to the usual bound and continuum states,
reflecting continuum states.
Moreover, a schematic plot of the spectrum as a function of $\theta_0$ reveals an energy gap 
region in the form of a triangle where no bound states can exist.
The end point of this region is a critical value $\theta_0^{\textrm{c}}=M/g$.
The zero 
mode is formed from the union of the threshold bound states present at this point and this zero mode exists for 
$\theta_0>\theta_0^{\textrm{c}}$.
For the J-R model $\theta_0^{\textrm{c}}=0$.
Hence the vacuum polarization is zero for $\theta_0\leqslant\theta_0^{\textrm{c}}$ 
and $\pm 1/2$ for $\theta_0>\theta_0^{\textrm{c}}$.
\par
In section 2 we define the massive J-R model which includes a massive fermion interacting 
with a prescribed background field in the form of the familiar kink. 
We then briefly discuss some important symmetries of our model
which are the same as the original massless J-R model.
In section 3 we obtain the second order decoupled Schr\"{o}dinger-like 
equations obtained from the first order Dirac equation.
Then, we solve these decoupled differential equations, analytically.
Depending on the range of energy, three kinds of states are possible for the fermion.
We first find the bound states in subsection 3.1.
The second kind of states which we call reflecting continuum states are obtained in subsection 3.2.
The wave functions of these states vanish at $x\to+\infty$, but are a superposition 
of an incident wave and a reflecting one with equal amplitude at $x\to-\infty$.
In subsection 3.3 we obtain and discuss the continuum scattering states.
The wave functions of these states are oscillatory at both spatial infinities.
In section 4 we plot the allowed energies of the fermion as a function 
of the parameters of the kink i.e.\;$\theta_0$ and $\mu$.
In these graphs we plot the energy levels of the bound states and also 
show the region for the energies of the reflecting and scattering states.
We observe that the zero mode in the massive J-R model is not always present 
and there is an energy gap in the form of a triangle in the $\theta_0$ graph in which no state is permitted.
In section 5 we summarize the results and draw some conclusions.
\section{The Model}
Consider a ($1+1$)-dimensional model including a Fermi field $\psi$ coupled to 
a pseudoscalar field $\phi_{\textrm{cl}}$, and defined by the following Lagrangian
\begin{equation}\label{lagrangian}\vspace{.2cm}
 {\cal L}=\bar{\psi}\left[i{\not}{\partial}-M -
g \phi_{\mathrm{cl}}(x)\right]\psi,
\end{equation}
where $M$ is the mass of the free fermion, $g$ is a positive coupling constant,  and $\phi_{\mathrm{cl}}(x)$ 
is a prescribed background field in the form of $\phi_{\mathrm{cl}}(x)=\big(m/\sqrt{\lambda}\big)\tanh\left(m x/\sqrt{2}\right)$
which is the kink of the $\phi^4$ theory.
Notice that the Lagrangian has an explicit fermion mass term and the mass of the fermion is nonzero even in the noninteracting case.
However, the interaction term changes the mass of the fermion at the tree level.
We can define two parameters $\theta_0=\frac{m}{\sqrt{\lambda}}$ and $\mu=\frac{m^2}{\sqrt{2\lambda}}$, 
which are the value of the kink at spatial infinity ($\phi(\infty)$) and its slope at $x=0$ 
($\frac{\mathrm{d}\phi}{\mathrm{d}x}\big|_{x=0}$), respectively.
We choose the following representation for the Dirac matrices: $\gamma^0=\sigma_1$ and $\gamma^1=i\sigma_3$.
This model possesses the charge conjugation symmetry.
This operator relates the states with positive energy to the ones with negative energy as 
$\psi^c_{-E}=\sigma_3\psi^*_E$ and a zero-energy fermionic mode, if it exists, is self charge conjugate, i.e.\;$\psi^c_{0}=\sigma_3\psi^*_0=\psi_0$.
One can easily check that this system also possesses the particle conjugation symmetry whose operation 
is $\psi_{-E}=\sigma_3\psi_E$.
Therefore, for every state with energy $E$, there is a corresponding state with energy 
$-E$ and the fermion spectrum is completely symmetric with respect to the line $E=0$.
This model is not invariant under the parity, since the background field is the kink which is an odd function in space.
Hence, this model does not preserve the CP and consequently it is not invariant under the time reversal.
Notice that all the symmetries of this massive model are the same as the massless one ($M=0$).
\par
In the following section we solve the equations of this model exactly and find the whole spectrum of the fermion, 
including the bound and continuum states.
\section{Spectrum of the fermion  in the presence of the background field}
\begin{center}
\begin{figure}[th] \hspace{4.cm}\includegraphics[width=7.5cm]{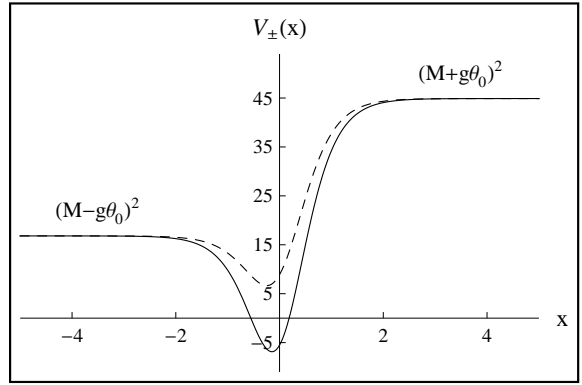}\caption{\label{potential} \small
The graphical representation of the potentials $V_{\pm}(x)$ as a function of $x$.
Solid and dashed lines show $V_{\pm}(x)$, respectively.
For this figure the parameters of the model are chosen to be $M=1.3$, $g=3.6$, $\theta_0=1.5$ and $\mu=2$.}
\end{figure}
\end{center}

The presence of the background filed can in general cause essential changes in the spectrum of the fermion. 
To find the spectrum of the fermion in our model, we solve the Dirac equation of the Lagrangian (\ref{lagrangian}).
Choosing $\psi(x,t)=\mathrm{e}^{-i Et}\left(\! \begin{array}{c}
\psi^{+}(x)\\
\psi^{-}(x)
\end{array}\! \right)$, the Dirac equation in the presence of the background field
$\phi_{\mathrm{cl}}(x)$ is as follows
\begin{equation}\label{eom}
\left(
\begin{matrix}
-\partial_x-M-g\phi_{\textmd{cl}}(x) & E \\
E & \partial_x-M-g\phi_{\textmd{cl}}(x)
\end{matrix}
\right)
\left(
\begin{matrix}
\psi^{+}(x) \\
\psi^{-}(x)
\end{matrix}
\right)=0.  
\end{equation}
This equation consists of two coupled first order differential equations.
In order to find the fermion spectrum, it is easier to first obtain the two 
decoupled second order equations obtained from Eq.\,(\ref{eom}).
Then, we can construct the solutions to the original Dirac Eq.\,(\ref{eom}).
The second order equations are two Schr\"{o}dinger-like equations which can be written as
\begin{align}\label{new2ndorder}
 \frac{\textmd{d}^2\psi^{\pm}(x)}{\textmd{d} x^2}+\left[\epsilon_{\pm}-V_{\pm}(x)\right]\psi^{\pm}(x)=0,
\end{align}
where $\epsilon_{\pm}=\left(E^{\pm}\right)^2$ and the potentials $V_{\pm}(x)$ are as follows
\begin{align}
V_{\pm}(x)=\left[ M +g\theta_0\tanh\left(\frac{\mu}{\theta_0}x\right)\right]^2 
\mp g\mu\,\mathrm{sech}^2\left(\frac{\mu}{\theta_0}x\right).
\end{align}
Figure \ref{potential} shows these potentials as a function of the spatial variable $x$ for a particular 
choice of the parameters of the model. 
For energies less than the asymptotic values of the potentials at $x=-\infty$
$\left(\left(M-g\theta_0\right)^2\right)$ and greater than the minima of the potentials, which are different 
for $V_{\pm}(x)$, some bound states with discrete energies are possible.
Also, all energies higher than $\left(M-g\theta_0\right)^2$ are allowed.
However, as we shall see, the continuum states with energies in the range 
$\left(M-g\theta_0\right)^2<\epsilon_{\pm}<\left(M+g\theta_0\right)^2$ are different 
from the continuum states with energies higher than $\left(M+g\theta_0\right)^2$.
From now on we use the redefinitions $x\to\frac{\theta_0}{\mu}x$, $g\to\frac{\mu}{\theta_0^2}g$, 
$E\to\frac{\mu}{\theta_0}E$ and $M\to\frac{\mu}{\theta_0}M$, for the brevity of the notation.
Applying these redefinitions, Eq.\,(\ref{new2ndorder}) remains the same and the potentials 
$V_{\pm}(x)$ change as follows
\begin{align}
V_{\pm}(x)=\left[ M +g\tanh\left(x\right)\right]^2 
\mp g\,\mathrm{sech}^2\left(x\right).
\end{align}
The solutions to Eq.\,(\ref{new2ndorder}), are well known \cite{morse,landau}.
Here we present a very short derivation of the solutions mainly for the purpose 
of setting up our notation.
We choose the following form for $\psi^{\pm}(x)$
\begin{align}\label{guess}
\psi^{\pm}(x)=\textmd{e}^{ -x a_{\pm}}\,\textmd{sech}^{b_{\pm}}\left(x\right)F_{\pm}(x).
\end{align} 
Substituting this ansatz into Eq.\,(\ref{new2ndorder}), we obtain
\begin{align}\label{subs}
\textmd{e}^{-x a_{\pm}}\textmd{sech}^{b_{\pm}}\left(x\right)
\bigg\{\frac{\textmd{d}^2F_{\pm}(x)}{\textmd{d}x^2}
-2\left[a_{\pm}+b_{\pm}~\textmd{tanh}\left(x\right)\right]\frac{\textmd{d}F_{\pm}(x)}{\textmd{d}x}
+\bigg[a_{\pm}^2+b_{\pm}^2-g^2+\left(E^{\pm}\right)^2-M^2\nonumber\\
+2\left(a_{\pm}b_{\pm}-g M\right)\tanh\left(x\right)
+\left(g^2\pm g-b_{\pm}(b_{\pm}+1)\right)\textmd{sech}^2\left(x\right)\bigg]F_{\pm}(x)\bigg \}=0.
\end{align}
We choose the following conditions
\begin{align}\label{conditions}
& a_{\pm}b_{\pm}-g M=0,\\\label{cond2}
&a_{\pm}^2+b_{\pm}^2-g^2+\left(E^{\pm}\right)^2-M^2=0,
\end{align}
which are equivalent to 
\begin{align}\label{aandb}
& a_{\pm}=\frac{1}{2}\left[\sqrt{\left(g+M\right)^2-\left(E^{\pm}\right)^2}
-\sqrt{\left(g-M\right)^2-\left(E^{\pm}\right)^2}\,\right]\equiv\frac{1}{2}(\kappa_1-\kappa_2),\\\label{aandb2}
&b_{\pm}=\frac{1}{2}\left[\sqrt{\left(g+M\right)^2-\left(E^{\pm}\right)^2}
+\sqrt{\left(g-M\right)^2-\left(E^{\pm}\right)^2}\,\right]\equiv\frac{1}{2}(\kappa_1+\kappa_2).
\end{align}
Using these conditions and an appropriate change of variables, 
i.e.\;$u=\frac{1}{2}\left[1-\textmd{tanh}\left(x\right)\right]$,  
which maps $x\in (-\infty,+\infty)$ to $u\in (1,0)$,
the differential equation of $F_{\pm}(x)$ turns
into a hypergeometric equation.
Therefore, the solution of Eq.\,(\ref{subs}) can be written as follows
\begin{align}\label{oursol}
&A~{}_{2}F_{1}\left(b_{\pm}+\frac{1}{2}-g\mp\frac{1}{2},b_{\pm}
+\frac{1}{2}+g\pm\frac{1}{2},1+a_{\pm}+b_{\pm}~;u\right)\nonumber\\
&+B~u^{-a_{\pm}-b_{\pm}}{}_{2}F_{1}\left(-a_{\pm}+\frac{1}{2}-g\mp\frac{1}{2}
,-a_{\pm}+\frac{1}{2}+g\pm\frac{1}{2},1-a_{\pm}-b_{\pm}~;u\right),
\end{align}
where ${}_{2}F_{1}\left(\alpha,\beta,\gamma;u\right)$ is the hypergeometric function, and A and B are the expansion coefficients.
These coefficients should be determined by the use of the asymptotic behavior and normalization 
of the wave functions.
\subsection{Bound states}
We devote this subsection to finding the bound state wave functions 
of the fermion and their associated discrete energies.
As we stated before, when $\epsilon_{\pm}<\left(M-g\right)^2$, some bound 
states are possible for the fermion and the equations of motion  would have 
solutions vanishing at spatial infinities. 
To find such solutions in which ~$\lim_{x\rightarrow \pm \infty} \psi^{\pm}(x)=0$,~
we set ~$b_{\pm}>a_{\pm}>0$~ to turn 
$\exp\big(-xa_{\pm}\big)\textmd{sech}^{b_{\pm}}\left(x\right)$ into a damping factor. 
The hypergeometric function is finite for $0<u<1$.
However, since both $b_{\pm}$ and $a_{\pm}$ are positive, 
$\lim_{u \to 0}u^{-a_{\pm}-b_{\pm}}=\infty$ and therefore  
the second term in the solution (\ref{oursol}) diverges. 
Setting $B=0$, the solution for our
equations would be as follows
\begin{align}\label{finaloursol}
\psi^{\pm}(x)&=\mathrm{e}^{-xa_{\pm}}\textmd{sech}^{b_{\pm}}\left(x\right)F_{\pm}(x)
=\frac{N_{\pm}\mathrm{e}^{-xa_{\pm}}}{\left(\textmd{e}^{x}+\textmd{e}^{-x}\right)^{b_{\pm}}}\nonumber\\
 &\times{}_{2}F_{1}\left(b_{\pm}+\frac{1}{2}-g\mp\frac{1}{2}
 ,b_{\pm}+\frac{1}{2}+g\pm\frac{1}{2},1+a_{\pm}+b_{\pm}~;
 \frac{\textmd{e}^{-x}}{\textmd{e}^{x}+\textmd{e}^{-x}}\right),
\end{align}
where $N_{\pm}$ are the normalization factors for the upper and lower components of the bound state wave functions, respectively.
These solutions have the proper behavior when $x\rightarrow +\infty$, 
i.e.\;$\lim_{x\to+\infty}\psi^{\pm}\propto \mathrm{e}^{-x(a_{\pm}+b_{\pm})}$. 
However, their behavior near $x=-\infty$ is as follows
\begin{align}\label{limit}
\lim_{x\to
-\infty}\psi^{\pm}(x)&=\frac{\Gamma(a_{\pm}+b_{\pm}+1)\Gamma(b_{\pm}-a_{\pm}){\textmd{e}^{x(a_{\pm}-b_{\pm})}}}
 {{\Gamma\left(b_{\pm}+\frac{1}{2}-g\mp\frac{1}{2}\right)}
 \Gamma\left(b_{\pm}+\frac{1}{2}+g\pm\frac{1}{2}\right)}\nonumber\\
&+\frac{\Gamma(a_{\pm}+b_{\pm}+1)\Gamma(a_{\pm}-b_{\pm})\textmd{e}^{x(b_{\pm}-a_{\pm})}}
 {\Gamma\left(a_{\pm}+\frac{1}{2}-g\mp\frac{1}{2}\right)\Gamma\left(a_{\pm}+\frac{1}{2}+g\pm\frac{1}{2}\right)}.
\end{align}
Since $b_{\pm}>a_{\pm}>0$, the first term in this equation diverges unless
the argument of one of the gamma functions in the denominator of this term is a semi-negative integer. 
Therefore, for the bound states we have the following constraint
\begin{align}\label{constraint}
b_{\pm}+\frac{1}{2}-g\mp\frac{1}{2}=-n,
\end{align}
where $n$ is a semi-positive integer. 
Using this constraint along with the constraints in Eqs.\,(\ref{conditions}, \ref{cond2}), 
the allowed discrete energies of the system are obtained.
These energies can be expressed in terms of the original parameters of the Lagrangian, i.e.\;$\theta_0$ and $\mu$, 
as follows 
\begin{align}\label{bnd}
\left(E_{n}^{+}\right)^2&={\left[1-\frac{M^2}{\left(g\theta_0-\frac{\mu}{\theta_0}n\right)^2}\right]\left(2g\mu n-\frac{\mu^2}{\theta_0^2}n^2\right)}
,~~~~n=0,1,2,\dots<\frac{\theta_0}{\mu}\left(g \theta_0-\sqrt{Mg\theta_0}\right),\\\label{bnd2}
\left(E_{n}^{-}\right)^2&={\left[1-\frac{M^2}{\left(g\theta_0-\frac{\mu}{\theta_0}n\right)^2}\right]\left(2g\mu n-\frac{\mu^2}{\theta_0^2}n^2\right)},
~~~~n=1,2,\dots<\frac{\theta_0}{\mu}\left(g \theta_0-\sqrt{Mg\theta_0}\right).
\end{align}
Notice that for the $n$th bound state to exist the parameters of the kink, 
i.e.\;$\theta_0$ 
and $\mu$, should satisfy the inequality 
$\left(\frac{g \theta_0^2}{\mu}-n\right)^2>\frac{M g \theta_0^3}{\mu^2}$ obtained from
 the condition $b_{\pm}>a_{\pm}$ and Eq.\,(\ref{constraint}).
Also, the upper bounds on the integer $n$ have been obtained using this relation.
The corresponding wave functions are
\begin{align}\label{waves}
\psi_{n}^{+}(x)& =\frac{N_{+}\textmd{e}^{-a\frac{\mu}{\theta_0}x}}
{\left(\textmd{e}^{\frac{\mu}{\theta_0}x}+\textmd{e}^{-\frac{\mu}{\theta_0}x}\right)^{\frac{g\theta_0^2}{\mu}-n}}
     ~{}_{2}F_{1}\left(-n,2\frac{g\theta_0^2}{\mu}-n+1,\frac{Mg\frac{\theta_0^2}{\mu}}{\frac{g\theta_0^2}{\mu}-n}+\frac{g\theta_0^2}{\mu}-n+1
     ;\frac{\textmd{e}^{-\frac{\mu}{\theta_0}x}}{\textmd{e}^{\frac{\mu}{\theta_0}x}
     +\textmd{e}^{-\frac{\mu}{\theta_0}x}}\right),\\\label{waves2}
     \psi_{n}^{-}(x)&
=\frac{N_{-}\textmd{e}^{-a\frac{\mu}{\theta_0}x}}{\left(\textmd{e}^{\frac{\mu}{\theta_0}x}+\textmd{e}^{-\frac{\mu}{\theta_0}x}
   \right)^{\frac{g\theta_0^2}{\mu}-n}}
~{}_{2}F_{1}\left(-n+1,2\frac{g\theta_0^2}{\mu}-n,\frac{Mg\frac{\theta_0^2}{\mu}}{\frac{g\theta_0^2}{\mu}-n}+\frac{g\theta_0^2}{\mu}-n+1
  ;\frac{\textmd{e}^{-\frac{\mu}{\theta_0}x}}{\textmd{e}^{\frac{\mu}{\theta_0}x}
  +\textmd{e}^{-\frac{\mu}{\theta_0}x}}\right),
\end{align}
where $a=\frac{M\frac{g\theta_0^2}{\mu}}{\frac{g\theta_0^2}{\mu}-n}$.
One can easily check that the solution $\psi_n(x)=\mathrm{e}^{-i E_nt}\left(\! 
\begin{array}{c}
\psi_n^{+}(x)\\
\psi_n^{-}(x)
\end{array}\! \right)
$ with $n=0,1,2,\dots$ satisfies the coupled first order Eq.\,(\ref{eom}), provided we set 
$N_-/N_+=\frac{n\mu}{\theta_0E_n}\left(1-\frac{M}{g\theta_0-\frac{\mu}{\theta_0}n}\right)$.
Notice that $\psi_n^{+}(x)$ and $\psi_n^{-}(x)$ in this doublet are the solutions to 
the second order equations, with the same energy $E_{n}=E_{n}^{+}=E_{n}^{-}$. 
\par
Now we focus our attention on the zero-energy mode (n=0). 
Notice that the energy of the lowest mode of the first Schr\"{o}dinger-like equation 
(the upper sign in Eq.\,(\ref{new2ndorder})) is zero.
However, the second equation (the lower sign in Eq.\,(\ref{new2ndorder})) does not have a zero-energy mode.
Hence, for the zero mode only $\psi_0^{+}(x)$ is nonzero; and we can easily obtain the 
explicit form of the spinor from Eq.\,(\ref{waves}).
However, it is useful to find it directly using the first order Dirac Eq.\,(\ref{eom}).
Setting $E=0$ in this equation, we obtain two decoupled first-order equations
which easily yield the following solutions
\begin{align}\label{zero}
\psi^+_0(x)=c_+~\textrm{e}^{-Mx} \left[\cosh\left(\frac{\mu}{\theta_0}x\right)\right]^{-\frac{g\theta_0^2}{\mu}},~~~~
\psi^-_0(x)=c_-~\textrm{e}^{Mx} \left[\cosh\left(\frac{\mu}{\theta_0}x\right)\right]^{\frac{g\theta_0^2}{\mu}},
\end{align}
where $c_+$ and $c_-$ are constant.
Since $\psi_0^-(x)$ makes the fermion wave function for the zero-energy mode unnormalizable, we set $c_-=0$.
Therefore, the wave function for this mode is as follows
\begin{equation}\label{zeromode}
\psi_0(x)=c_+
\left(
\begin{matrix}
 \textrm{e}^{-Mx} \left[\cosh\left(\frac{\mu}{\theta_0}x\right)\right]^{-\frac{g\theta_0^2}{\mu}}\vspace{0.2cm}\\
0
\end{matrix}
\right).
\end{equation}
One can easily see that we should have $\theta_0>\theta_0^{c}=M/g$ to have a normalizable bound state. 
Notice that at $\theta_0=\theta_0^{c}$ there exist two half-bound states, whose
wave functions approach nonzero constants at spatial infinities, and just after that 
the zero-energy bound state is formed from the union of these threshold bound states and continues to exist for $\theta_0>\theta_0^{c}$.
\subsection{Continuum reflecting states}
Suppose that the energy $\epsilon_{\pm}$ is greater than $(M-g\theta_0)^2$ but 
smaller than $(M+g\theta_0)^2$ (see Fig.\,{\ref{potential}}).
In this range the quantity $\kappa_1=\frac{\theta_0}{\mu}\sqrt{(g\theta_0+M)^2-(E_{\pm})^2}$ 
is real but the quantity $\kappa_2=\frac{\theta_0}{\mu}\sqrt{(g\theta_0-M)^2-(E_{\pm})^2}$ $=-ik_2$ 
is imaginary (see Eqs.\,({\ref{aandb},\ref{aandb2}})).
As we know, continuum states are possible in this range of energy.
The wave functions of these states vanish when $x\rightarrow +\infty$.
However, they are oscillatory at $x\rightarrow -\infty$.
Since when $x\rightarrow +\infty$, the first solution in Eq.\,(\ref{oursol}) behaves as 
$\mathrm{e}^{-(a_{\pm}+b_{\pm})\frac{\mu x}{\theta_0}}$,
$a_{\pm}+b_{\pm}=2\kappa_1$ should be positive.
However, the second solution in Eq.\,(\ref{oursol}) does not have the proper behavior when 
$x\rightarrow +\infty$ and we should set again $B=0$.
Thus, the wave functions of these solutions are as follows
\begin{align}\label{reflectwave}
\psi_{\mathrm{crs}}^{\pm}(x)&=
\frac{N^{\mathrm{crs}}_{\pm}\mathrm{e}^{-\frac{1}{2}(\kappa_1+ik_2)\frac{\mu}{\theta_0}x}}
{\left(\textmd{e}^{\frac{\mu}{\theta_0}x}+\textmd{e}^{-\frac{\mu}{\theta_0}x}\right)^{\frac{1}{2}\left(\kappa_1-ik_2\right)}}\nonumber\\
     &\times
{}_{2}F_{1}\left(\frac{1}{2}\left(\kappa_1-ik_2\right)+\frac{1}{2}-\zeta_{\pm},\frac{1}{2}\left(\kappa_1-ik_2\right)+\frac{1}{2}+\zeta_{\pm},
\kappa_1+1;
\frac{\textmd{e}^{-\frac{\mu}{\theta_0}x}}{\textmd{e}^{\frac{\mu}{\theta_0}x}
     +\textmd{e}^{-\frac{\mu}{\theta_0}x}}\right),
\end{align}
where $\zeta_{\pm}=\frac{g\theta_0^2}{\mu}\pm\frac{1}{2}$ and $N_{\pm}^{\mathrm{crs}}$ are the normalization factors for these states.
We can easily check that these wave functions satisfy Eq.\,(\ref{eom}) if we set 
$N^{\mathrm{crs}}_-/N^{\mathrm{crs}}_+
=(-\frac{\mu}{\theta_0}\kappa_1+g\theta_0+M)
/\sqrt{-\frac{\mu^2}{\theta^2_0}\kappa_1^2+(g\theta_0+M)^2}$.
These solutions behave as $\mathrm{e}^{-\kappa_1\frac{\mu}{\theta_0}x}$ when $x\rightarrow +\infty$ and their asymptotic behavior at the other boundary, 
i.e.\;$x\rightarrow -\infty$, is as follows
 \begin{align}\label{reflectasymp}
\psi_{\mathrm{crs}}^{\pm}(x\rightarrow -\infty)=
 N_{\pm}^{\mathrm{crs}} \Gamma(1+\kappa_1)
\bigg[
&\frac{\Gamma(-ik_2)~\mathrm{e}^{ik_2\frac{\mu}{\theta_0}x}} 
{\Gamma\left(\frac{1}{2}+\zeta_{\pm}+\frac{\kappa_1}{2}-\frac{ik_2}{2}\right)
 \Gamma\left(\frac{1}{2}-\zeta_{\pm}+\frac{\kappa_1}{2}-\frac{ik_2}{2}\right)}
\nonumber\\
&+\frac{\Gamma(ik_2)~\mathrm{e}^{-ik_2\frac{\mu}{\theta_0}x}}{\Gamma\left(\frac{1}{2}+\zeta_{\pm}+\frac{\kappa_1}{2}+\frac{ik_2}{2}\right)
 \Gamma\left(\frac{1}{2}-\zeta_{\pm}+\frac{\kappa_1}{2}+\frac{ik_2}{2}\right)}
\bigg].
\end{align}
The first term represents an incident wave at $x=-\infty$, traveling in the positive direction 
($\mathrm{e}^{ik_2\frac{\mu}{\theta_0}x}$), and the second term a reflected wave at $x=-\infty$,  
travelling in the negative direction ($\mathrm{e}^{-ik_2\frac{\mu}{\theta_0}x}$).
\subsection{Continuum scattering states} 
Now, we focus our attention on the states with energies greater than $(M+g\theta_0)^2$. 
In this range  both the quantities $\kappa_1=-ik_1$ and $\kappa_2=-ik_2$ are imaginary.
All the energies of this range are permitted and we have continuum states.
The wave functions of these states should be oscillatory in both of the spatial infinities, 
i.e.\;$x\rightarrow \pm \infty$, and are as follows
\begin{align}\label{scatt}
\psi_{\mathrm{css}}^{\pm,{\mathrm{L}}}(x)&=\frac{N_{\pm,{\mathrm{L}}}^{\mathrm{css}}\mathrm{e}^{\frac{i}{2}(k_1-k_2)\frac{\mu}{\theta_0}x}}{\left(\mathrm{e}^{\frac{\mu}{\theta_0}x}+\mathrm{e}^{-\frac{\mu}{\theta_0}x}\right)^{\frac{-i}{2}(k_1+k_2)}}
\nonumber\\&\times
{}_{2}F_1\left(\frac{-i}{2}(k_1+k_2)+\frac{1}{2}-\zeta_{\pm}, \frac{-i}{2}(k_1+k_2)+\frac{1}{2}+\zeta_{\pm} ,1-ik_1;\frac{\mathrm{e}^{-\frac{\mu}{\theta_0}x}}{\mathrm{e}^{\frac{\mu}{\theta_0}x}
+\mathrm{e}^{-\frac{\mu}{\theta_0}x}}\right),
\end{align}
\begin{align}\label{scatt2}
\psi_{\mathrm{css}}^{\pm,{\mathrm{R}}}(x)&=
\frac{N_{\pm,{\mathrm{R}}}^{\mathrm{css}}\mathrm{e}^{\frac{i}{2}(k_1-k_2)\frac{\mu}{\theta_0}x}}{\left(\mathrm{e}^{\frac{\mu}{\theta_0}x}+\mathrm{e}^{-\frac{\mu}{\theta_0}x}\right)^{\frac{-i}{2}(k_1+k_2)}}
\nonumber\\&\times
{}_{2}F_1\left(\frac{-i}{2}(k_1+k_2)+\frac{1}{2}-\zeta_{\pm}, \frac{-i}{2}(k_1+k_2)+\frac{1}{2}+\zeta_{\pm} ,1-ik_2;\frac{\mathrm{e}^{\frac{\mu}{\theta_0}x}}{\mathrm{e}^{\frac{\mu}{\theta_0}x}
+\mathrm{e}^{-\frac{\mu}{\theta_0}x}}\right),
\end{align}
where $N_{\pm,{\mathrm{L}}}^{\mathrm{css}}$ and 
$N_{\pm,{\mathrm{R}}}^{\mathrm{css}}$ are the normalization factors for the continuum scattering states.
These solutions satisfy Eq.\,(\ref{eom}), when 
$N^{\mathrm{css}}_{-,{\mathrm{L}}}/N^{\mathrm{css}}_{+,{\mathrm{L}}}
=(i\frac{\mu}{\theta_0}k_1+g\theta_0+M)
/\sqrt{\frac{\mu^2}{\theta^2_0}k_1^2+(g\theta_0+M)^2}$
and 
$N^{\mathrm{css}}_{-,{\mathrm{R}}}/N^{\mathrm{css}}_{+,{\mathrm{R}}}
=(-i\frac{\mu}{\theta_0}k_2-g\theta_0+M)
/\sqrt{\frac{\mu^2}{\theta^2_0}k_1^2+(g\theta_0+M)^2}$.
The asymptotic behavior of these wave functions at the spatial infinities is as follows
\begin{align}\label{asympscatt}
\psi^{\pm,{\mathrm{L}}}_{\mathrm{css}}=\begin{cases}
 N_{\pm,{\mathrm{L}}}^{\mathrm{css}} 
\bigg[\frac{\Gamma(1-ik_1)\Gamma(-ik_2)~\mathrm{e}^{ik_2\frac{\mu}{\theta_0}x}}
{\Gamma\left(\frac{1}{2}+\zeta_{\pm}-\frac{ik_1}{2}-\frac{ik_2}{2}\right)
 \Gamma\left(\frac{1}{2}-\zeta_{\pm}-\frac{ik_1}{2}-\frac{ik_2}{2}\right)}
 +
 \frac{\Gamma(1-ik_1)\Gamma(ik_2)~\mathrm{e}^{-ik_2\frac{\mu}{\theta_0}x}}
{\Gamma\left(\frac{1}{2}+\zeta_{\pm}-\frac{ik_1}{2}+\frac{ik_2}{2}\right)
 \Gamma\left(\frac{1}{2}-\zeta_{\pm}-\frac{ik_1}{2}+\frac{ik_2}{2}\right)}\bigg],
 &\mathrm{as}~x\rightarrow-\infty,\vspace{.3cm}\\
 N_{\pm,{\mathrm{L}}}^{\mathrm{css}}~\mathrm{e}^{ik_1\frac{\mu}{\theta_0}x},&\mathrm{as}~x\rightarrow+\infty,
\end{cases}
\end{align}
\begin{align}\label{asympscatt2}
\psi^{\pm,{\mathrm{R}}}_{\mathrm{css}}=\begin{cases}
 N_{\pm,{\mathrm{R}}}^{\mathrm{css}}~\mathrm{e}^{-ik_2\frac{\mu}{\theta_0}x},&\mathrm{as}~x\rightarrow-\infty,\vspace{.3cm}\\
 N_{\pm,{\mathrm{R}}}^{\mathrm{css}}
 \bigg[\frac{\Gamma(1-ik_2)\Gamma(ik_1)~\mathrm{e}^{ik_1\frac{\mu}{\theta_0}x}}
 {\Gamma\left(\frac{1}{2}+\zeta_{\pm}+\frac{ik_1}{2}-\frac{ik_2}{2}\right)
 \Gamma\left(\frac{1}{2}-\zeta_{\pm}+\frac{ik_1}{2}-\frac{ik_2}{2}\right)}
 +
 \frac{\Gamma(1-ik_2)\Gamma(-ik_1)~\mathrm{e}^{-ik_1\frac{\mu}{\theta_0}x}}
 {\Gamma\left(\frac{1}{2}+\zeta_{\pm}-\frac{ik_1}{2}-\frac{ik_2}{2}\right)
 \Gamma\left(\frac{1}{2}-\zeta_{\pm}-\frac{ik_1}{2}-\frac{ik_2}{2}\right)}\bigg],
 &\mathrm{as}~x\rightarrow+\infty.
\end{cases}
\end{align}
The asymptotic behavior of $\psi^{\pm,{\mathrm{L}}}_{\mathrm{css}}(x)$ (Eq.\,(\ref{scatt})), shown in Eq.\,(\ref{asympscatt}), corresponds to an incident wave at $x\rightarrow -\infty$ moving to the right 
($\mathrm{e}^{ik_2\frac{\mu}{\theta_0}x}$), a reflected wave at $x\rightarrow -\infty$ moving back to the left ($\mathrm{e}^{-ik_2\frac{\mu}{\theta_0}x}$) 
and a transmitted wave at $x\rightarrow +\infty$ moving to the right ($\mathrm{e}^{ik_1\frac{\mu}{\theta_0}x}$).
We can refer to this as a left-scattering process.
Analogously, the asymptotic behavior of $\psi^{\pm,{\mathrm{R}}}_{\mathrm{css}}(x)$ (Eq.\,(\ref{scatt2})), 
shown in Eq.\,(\ref{asympscatt2}), describes a right-scattering process.
\section{Graphical representation of the fermion spectrum}  
\begin{center}
\begin{figure}[th] \hspace{1.3cm}\includegraphics[width=13.cm]{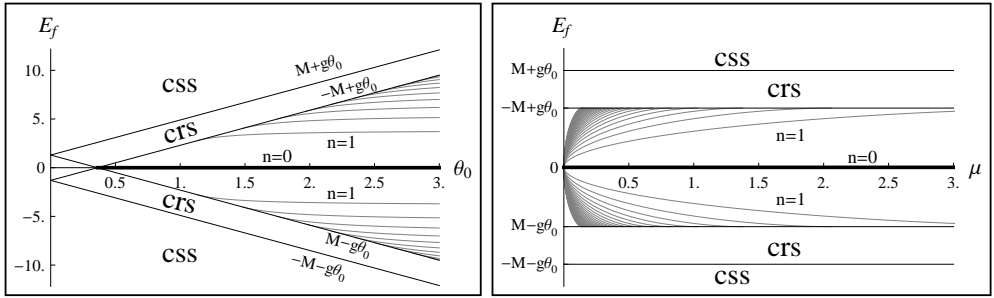}\caption{\label{bnd1} \small
The left graph shows the energies of the fermion 
as a function of $\theta_0$ at $\mu=2$
and the right graph shows the energies as a function of $\mu$ at $\theta_0=1.5$ for $n\leqslant 30$.
In both graphs $M=1.3$ and $g=3.6$.
In these graphs we show the zero-energy fermionic mode with two bold lines.
For this mode $n=0$.}
\end{figure}
\end{center}

\begin{center}
\begin{figure}[th] \hspace{2.2cm}\includegraphics[width=11.cm]{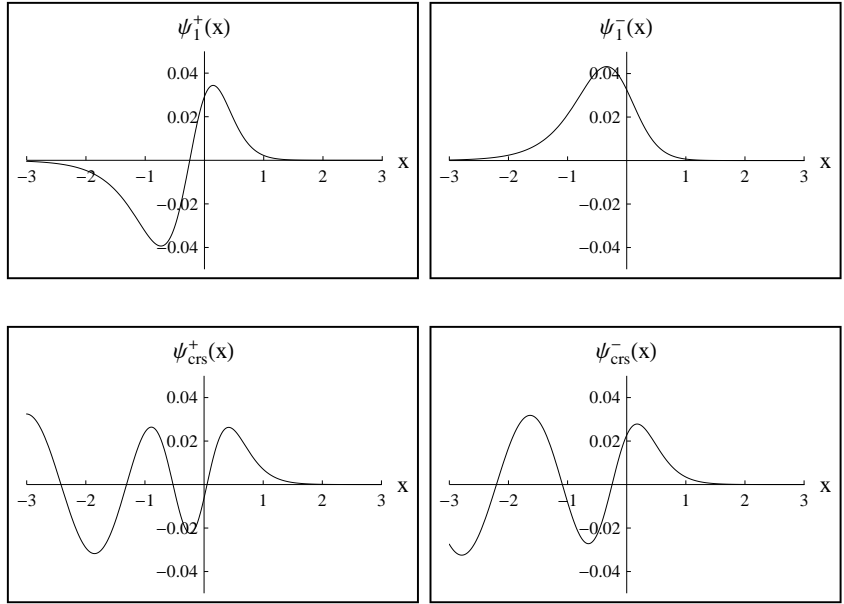}\caption{\label{wavef} \small
The upper two graphs show the upper and lower components of the fermion wave function for the bound state with $n=1$ as a function of the spatial variable $x$. 
The lower graphs show the upper and lower components of a sample of the reflecting continuum states.
The parameters of the model in all graphs are $M=1.3$, $g=3.6$, $\mu=2$ and $\theta_0=1.5$.}
\end{figure}
\end{center}

In this section we show the energies of the fermion in some graphs.
In the left and right graphs of Fig.\,\ref{bnd1} we depict the bound state energies as a function of the parameters $\theta_0$ and $\mu$, respectively. 
This figure also shows the energies of the continuum reflecting and scattering states, denoted by \lq{}crs\rq{} and \lq{}css\rq{}, respectively.
The zero-energy bound state is shown with a bold line in these graphs.
As is well known, the zero-energy mode in the J-R model, which is the origin of the fractional fermion number $\pm 1/2$ for the ground state, 
is always present, independent of the parameters of the model.
The free fermion in this model has no explicit mass term.
Therefore, there is no mass gap for the free Dirac field and the two threshold half-bound states present for the free case in ($1+1$) dimensions have both zero energy in J-R model.
However, by turning up the potential, a mass gap appears and the two zero-energy 
half-bound states merge to form the single zero-energy bound state.
However, the situation is different for the massive J-R model.
As can be seen in the left graph of Fig.\,\ref{bnd1}, there exists a mass gap for the free fermion of the massive J-R model and the energies of the two threshold half-bound states in the zero strength of the potential are $\pm M$.
By increasing the value of $\theta_0$, these two states continue being threshold bound states with energies $\pm M_{\textrm{f}}=\pm (M-g\theta_0)$ until 
the lines of $\pm M_{\textrm{f}}$ cross each other and become zero at $\theta_0^{\textrm{c}}=M/g$.
After this point the two threshold bound states form a zero-energy bound state.
This zero mode is present for $\theta_0$ greater than $M/g$ and therefore from this point on the fermion number of the vacuum becomes $\pm 1/2$ as in the J-R model.
In addition to this mode, some other fermionic bound states separate from the lines $\pm M_{\textrm{f}}$ for $\theta_0>M/g$.
Notice that no bound states exist in the triangular region.
In Fig.\,\ref{wavef} we show some samples of the wave functions of the bound states and the continuum reflecting states.
As can be seen, all these graphs have the proper asymptotic behavior.  

\section{Conclusion}
In this paper we introduce and thoroughly investigate a massive Jackiw-Rebbi model 
containing a massive fermion coupled to a prescribed background field in the form 
of the kink.
The only difference between this model and the original J-R model is that in the present model the fermion has a mass term even in the zero strength of the potential.
We solve the equations of this model exactly and analytically, for arbitrary choice of the parameters of the kink, and find the whole spectrum of the interacting fermion.
We show the energies of all the states of the fermion, including the discrete bound states, the continuum reflecting states and the continuum scattering states and some samples of the wave functions in some graphs.
We find that the mechanism of dynamical mass generation is common to both models.
In the graph of the energies of the fermion as a function of $\theta_0$ we see an energy region in the form of a triangle, 
in which no bound state for the fermion is allowed including the zero mode.
The zero-energy bound state exists only for $\theta_0>\theta_0^{\textrm{c}}=M/g$ and 
this is in sharp contrast to the original J-R model where the zero mode is always present, 
regardless of the values of the parameters of the model.
Consequently, the kink in the massive J-R model does not always polarize the vacuum 
and vacuum polarization jumps between the value zero and $\pm 1/2$ at 
$\theta_0=\theta_0^{\textrm{c}}$.
\section*{Acknowledgement} We would like to thank the research office
of the Shahid Beheshti University for financial support.


 \end{document}